\documentclass[aps,epsf,amssymb,latexsym,showpacs,amsmath,preprint]{revtex4}
\usepackage[latin1]{inputenc}
\usepackage{graphics}
\usepackage{amsfonts, amssymb, latexsym, amsmath}
\input epsf.sty

\begin{document}
\title{Effect of confinement on coil-globule transition\\
}
\author{P. K. Mishra$^1$ and Sanjay Kumar$^{1,2}$}
%\author{a$^1$, b$^{1,2}$ and c$^1$}
\email{pramod@justice.com (P. K. Mishra); yashankit@yahoo.com (S. Kumar)}
\affiliation{ $^1$Department of Physics\\
Banaras Hindu University,\\
Varanasi-221 005, India\\
$^2$Max-Plank Institute for Complex System \\
Nothnitzer Streat 38, 01187 Dresden, Germany}
\begin{abstract}
The equilibrium thermodynamic properties of a linear polymer chain
confined to a space between two impenetrable walls (lines) at a distance 
$D$ under various solvent conditions have been studied  using series
analysis and exact enumeration technique. We have calculated the end to
end distance of polymer chain, which shows a non-monotonic behaviour with inter wall 
separation $D$. The density distribution profile shows a maxima at 
a particular value of ${(D=)}D^*$. Around this $D^*$, our results show that the collapse
transition occurs at higher temperature as compared to its bulk value 
of $2d$ and $3d$. The variation of $\theta-$ temperature with $D$ shows
a re-entrance behaviour. We also calculate the force of compression 
exerted by the walls (lines) on the polymer.
\end{abstract}
\maketitle
\section{Introduction}
In last few years$^{1-6}$ much 
attention has been paid on the subject of confinement of polymer and its scaling
relations. This is due to the fact that a
polymer chain in restricted geometries exhibits specific and interesting properties which
find application in steric stabilization of colloidal dispersions, 
thin films, adsorption behaviour of gels, surface coatings, membrane in nano pores etc$^{6,7}$.
Several theoretical and numerical attempts has been made to understand these properties$^{1,6,10}$ 
A polymer chain is said to be confined when the spacing $D$ between the parallel walls 
(lines in $2d$) is less than the average end to end distance $<R_E>$ of the chain. The parallel ($<R_{E||}>$)and 
perpendicular component ($<R_{E\perp}>$) of $<R_E>$ is expected to behave as$^2$,in $3d$, 
\begin{eqnarray}
<R_{E||}>\sim a.N^{\frac{3}{4}} {(\frac{D}{a}})^{1-\frac{3}{4\nu}}
\label{@mome}\end{eqnarray}
and 
\begin{eqnarray}
<R_{E\perp}>\sim D.
\label{@}\end{eqnarray}
Where $N$ is the total number of monomers in the chain, $a$ is the lattice parameter 
and $\nu$ is the end to end distance exponent respectively. When $D$ is greater 
than $ <R_E>$, the effect of confinement on its conformational properties vanishes i.e.
\begin{eqnarray}
<R_{E||}> = <R_{E\perp}> \sim a N^{\nu}
\label{@}\end{eqnarray}
The monomer density ($\rho(z)$) profile near the wall is expected to behave as 
\begin{eqnarray}
\rho(z)\sim z^{\frac{1}{\nu}} 
\label{@}\end{eqnarray}
here $z$ is the perpendicular distance from the surface. It is assumed
that the force exerted by the polymer on the wall is proportional to the monomer
density near the wall. It has been shown that$^9$
\begin{eqnarray}
{\lim}_{z\to 0}k\frac{\rho(z)}{z^{\frac{1}{\nu}}}\sim B \frac{f}{k_BT} 
\label{@}\end{eqnarray}
where $k$ is the non-universal amplitude which relates the end to end distance
$<R_E>$ of a polymer to the chain length $N$, $B$ is the universal amplitude
ratio, $f$ is the force exerted on the wall due to confinement of the
chain and $T$ is the temperature of the system respectively.
Considerable efforts have been made$^{1-10}$ to verify 
the scaling relations and determine the value of $B$ in two and three 
dimensions. In a recent simulation by Hsu and Grassberger$^1$ much
better value of $B$(=$2.04\pm 0.04$ (in $2d$) and $2.13\pm 0.11$  (in $3d$)) has been 
obtained as compared to the previous Monte Carlo estimates$^{2,10}$.

The complete phase diagram of surface interacting polymer chain
in poor solvent condition has been studied recently$^{11,12}$.
Our aim in this paper is to analyze the effect of geometrical confinement such as
the one given by a slit(or strip) on coil-globule transition under 
%various solvent
%{\it various solvent}
good and poor solvent 
conditions. In any Monte Carlo simulation, and particularly in the study
of polymeric system, improper sampling of hyper-space under finite number
of steps being considered which some time overlook some of the interesting
features of these systems$^{11,13}$. Keeping this in mind, we 
have analyzed the partition function through
series analysis and exact enumeration technique. We prefer
this technique because in this case the complete partition function
can be analyzed exactly and the scaling corrections are
correctly taken into account by a suitable extrapolation scheme$^{13,14}$. 
For the sake of simplicity and comparison with the previous results 
the walls here have been considered as neutral.

The work is organized as follows; In Section $2$, we briefly describe 
the model 
and the method. Section $3$ deals with the monomer density profile in two and three 
dimensions where we compare our results with the known results. In Section $4$ 
we study the effect of confinement on the coil-globule transition and establish
the phase diagram which  shows the reentrance behaviour in 
three dimensions. The paper ends with the discussion in Section $5$. 
\section{MODEL AND METHOD}
We represent a polymer chain by self-avoiding  walks (SAWs) on a square ($2d$) and cubic
($3d$) lattices. The polymer chain is confined in between strip
of two impenetrable surface of height $D =0,1,2 ..16$ in case of square lattice and
$D =0,1,2 ..8$ for cubic lattice, respectively. The walk originates from
the middle of the surface (kept fixed) and spreads along all possible
directions. This ensures that the chain
is not affected by the presence of the upper surface for large values of D and 
one recovers the usual statistics of SAWs. The thermodynamic properties
associated with the confined polymer may be obtained from the partition 
function which can be written as a sum over all possible configurations 
confined in distance $D$ i.e.
\begin{eqnarray}
Z_N (D) = \sum_{C_N} C_{N}(D) x^{N}
\label{@}\end{eqnarray}
here $x$ is the fugacity associated with each step of the walk and  $C_{N}(D)$ is 
the number distinct $SAWs$ of $N$ steps confined in width $D$. 
We have obtained $C_{N}(D)$ up to $N \le 28$  
in two dimensions (for $D \le 16$) and $N \le 19$  (for $ D \le 8$)
in three dimensions. In general, it is appropriate to assume that for large $N$,
\begin{eqnarray}
C_{N}(D) \sim N^{\gamma - 1} \mu(D)^{N}
\label{@}\end{eqnarray}
here $\mu(D)$ is the connectivity constant of the lattice and $\gamma$ is
the associated critical exponent. The value of $\mu(D)$ can be estimated
using the ratio method$^{13-14}$ with an associated Neville table. These values are
found to be in good agreement with the result obtained by the exact
transfer matrix calculation$^{15}$. The end to end distance
exponent $\nu$ may be found using the relation 
\begin{eqnarray}
\nu= \frac{\log(<R_{N}>/<R_{N-2}>)}{\log (N/N-2)}
\label{@}\end{eqnarray}
where $<R_N>$ is the end to end distance exponent of $N$ steps walk. The values
of $\nu=0.748\pm 0.002(2d)$ and $0.58\pm0.01(3d)$ are in good agreement with the known
results in the limit $D\to \infty^6$. For finite value of $D$, we restore
the scaling proposed by Milchev and Binder$^2$. The slopes
found for $2d$ and $3d$ for
parallel and perpendicular component of end to end distance are in 
good agreement with Eqs. ($1-2$) as shown in Figs. ($1-2$). 
\begin{figure}[htp!]
\begin{center}
\setlength{\unitlength}{1cm}
\begin{picture}(5,5)
\put(-4,-.1){\scalebox{.5}[.5]{\includegraphics*{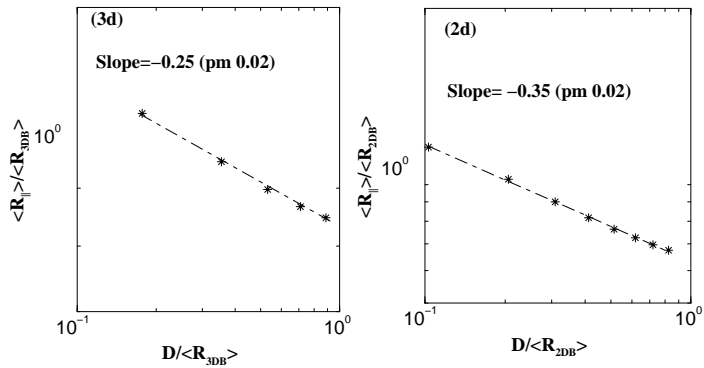}}}
\end{picture}
\caption[1a]{Scaling for parallel component of end to end
 distance exponent. }
\label{1a}
\end{center}
\end{figure}
\begin{figure}[htp!]
\begin{center}
\setlength{\unitlength}{1cm}
\begin{picture}(5,5)
\put(-4,-.1){\scalebox{.5}[.5]{\includegraphics*{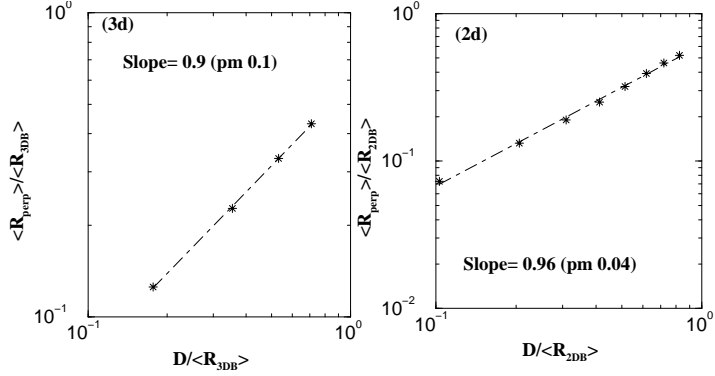}}}
\end{picture}
\caption[1b]{Scaling for perpendicular component of end to
 end distance.}
\label{1b}
\end{center}
\end{figure}
However, the variation $<R_E>$ is non-monotonic with $D$
as shown in Fig. $3$. We find that $<R_E>$  first decreases with decreasing $D$
before it starts to rise. This behaviour was ascribed earlier to a lateral
squeezing of chains with increasing $D$, followed by an elongation in the
direction parallel to the walls$^{16-17}$
\begin{figure}[htp!]
\begin{center}
\setlength{\unitlength}{1cm}
\begin{picture}(5,5)
\put(-4,-.1){\scalebox{.5}[.5]{\includegraphics*{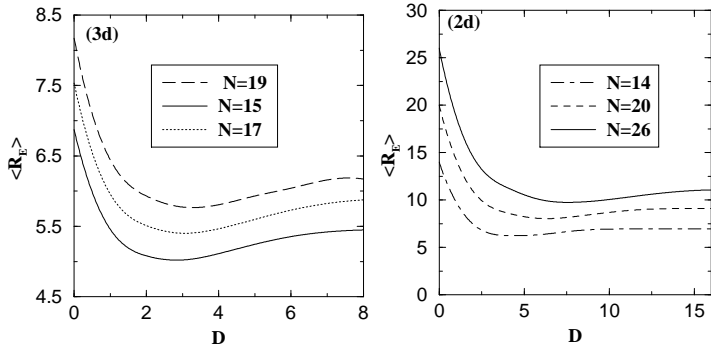}}}
\end{picture}
~\\
~\\[-1em]
\caption[1c]{The variation of end to end distance ($<R_E>$) of
  polymer chain for various length with inter wall
  separation $D$ for $3d$ and $2d$. }
\label{1c}
\end{center}
\end{figure}
\section{Monomer density distributions}
In the present model, one end of the chain is anchored at the middle
of one of the walls. Consequently, it introduces asymmetry in the distribution
of monomer density in the direction perpendicular to the plane of
wall. In earlier simulations$^1$, the walker has been allowed to start from
any of the planes in between the walls, therefore, the monomer distribution in such 
case is found to be symmetric. The density distribution of polymer
attached to the surface has been shown in Fig. $4$ for $2d$ and $3d$. From these
figures it is obvious that when  the thickness of the strip decreases, the monomer 
density near the wall increases. In Fig. $5$, we show the monomer density
for fixed $D$ which increases with $N$,and therefore, is not a finite size effect. 
In fact the peak shifts 
%{\it out to larger $D$} {\bf 
away from surface, at which polymer is grafted, as $N$ increases.

\begin{figure}[htp!]
\begin{center}
\setlength{\unitlength}{1cm}
\begin{picture}(9,9)
\put(-4,-.1){\scalebox{.9}[.9]{\includegraphics*{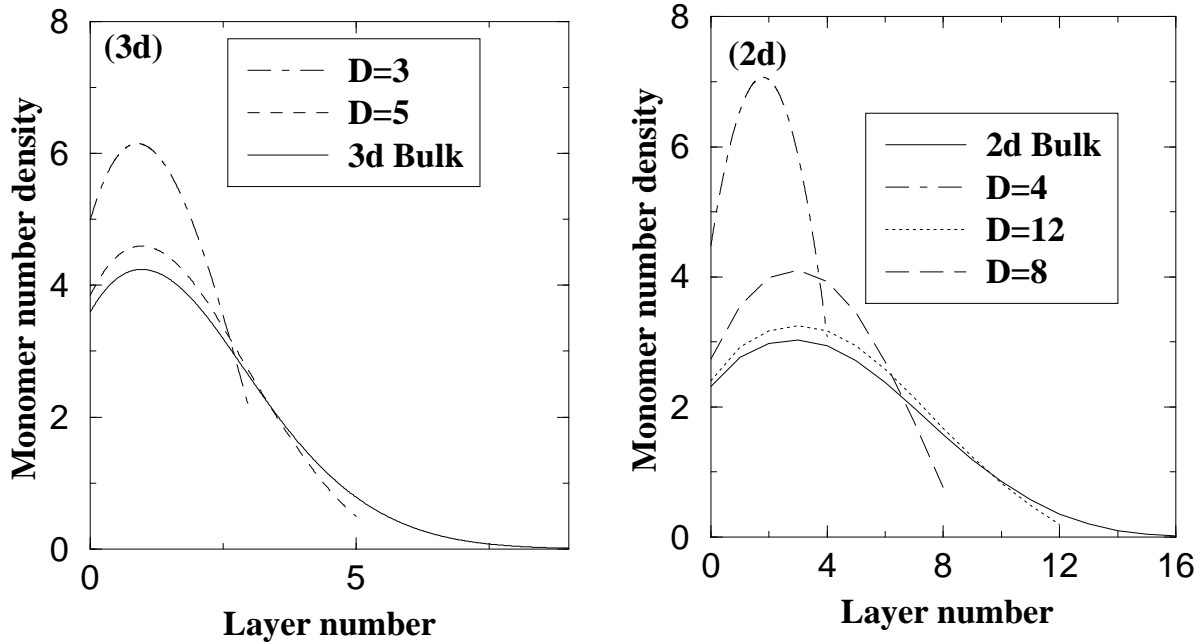}}}
\end{picture}
\caption[2a]{ The variation of monomer density with $D$ for different value of $D$. }
\label{2a}
\end{center}
\end{figure}
The force exerted on the wall is 
proportional to the monomer density on the surface and may be expressed in 
terms of the work done by moving one of the surfaces

\begin{eqnarray}
f=-\frac{\partial G}{\partial D}=\frac{k_B T}{\mu } \frac{\partial \mu}{\partial D}.
\label{@}\end{eqnarray}

Using the simple scaling ansatz 
\begin{eqnarray}
\rho(z)=\frac{1}{D+1} f_\rho(\xi)=\frac{1}{D+1} A\xi (1-\xi )^{1/\nu_d}
\label{@}\end{eqnarray}

where $\xi=y/D+1$ and $\nu_d$ is the end to end distance 
exponent of 
$d$ dimension, with $A=10.38(2d)$ and $18.74(3d)$, 
as proposed by Hsu and Grassberger$^1$
we find the value of the universal
amplitude ratio B to be equal to $2.01(2d)$ and $2.1(3d)$, respectively. 
%{\bf Should we include here final expression that we are using for $B$ and write
%value of $k$, ${\mu}$, $\nu$ and $a$ for two and three dimensions. }
These values are
in good agreement with the exact result$^9$ and simulations$^1$.
\begin{figure}[htp!]
\begin{center}
\setlength{\unitlength}{1cm}
\begin{picture}(9,9)
\put(-4,-.1){\scalebox{.7}[.7]{\includegraphics*{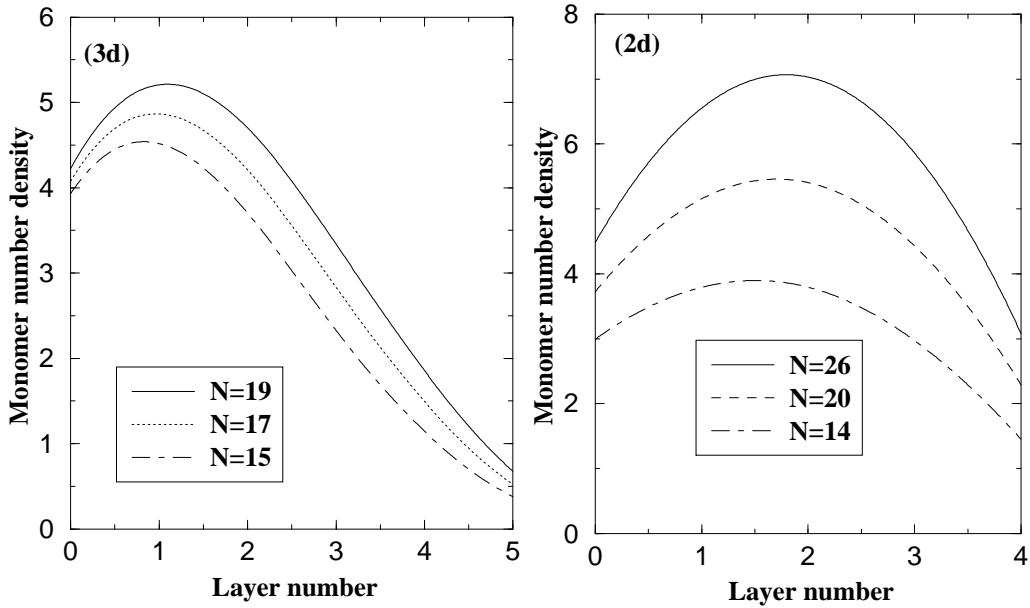}}}
\end{picture}
\caption[2b]{The density profile with $D$ for different chain
  length. These figures show that the peak height increases with $N$ but shifts 
out to larger $D$ as $N$ increases.  }
\label{2b}
\end{center}
\end{figure}
\section{Effect of confinement on Coil-Globule phase transition }
As confinement affects the monomer density profile, one expects 
a significant change in the transition point if one varies the
 distance between the two walls. A square lattice with finite $D$ is
essentially an one dimensional system and hence there is no phase 
transition as such. Keeping this in mind, we  study such an effect,
on a simple cubic lattice with variable wall thickness $D$.
We consider self- attracting-self avoiding walks (SASAWs) and 
associate energy $-\epsilon_p$ for nearest
neighbour monomers which are not chemically bonded. The partition function
of the confined chain may be written as
\begin{eqnarray}
Z_N (D,u) = \sum_{N_p} C_{N,N_p}(D) u^{N_p}  
\label{@}\end{eqnarray}
where $u$ is the Boltzmann weight defined as $u=e^{- \epsilon_p/{k_\beta}T}$ 
and $N_p$ is the number 
of pairs of monomer of chain length $N$. $C_{N,N_p}(D)$ is the number of distinct
configurations with $N_p$ number of nearest neighbour pairs confined 
between the walls at distance $D$. 
The reduced free energy can be written as	
\begin{figure}[htp!]
\begin{center}
\setlength{\unitlength}{1cm}
\begin{picture}(7,10)
\put(-1.8,-.1){\scalebox{.9}[.9]{\includegraphics*{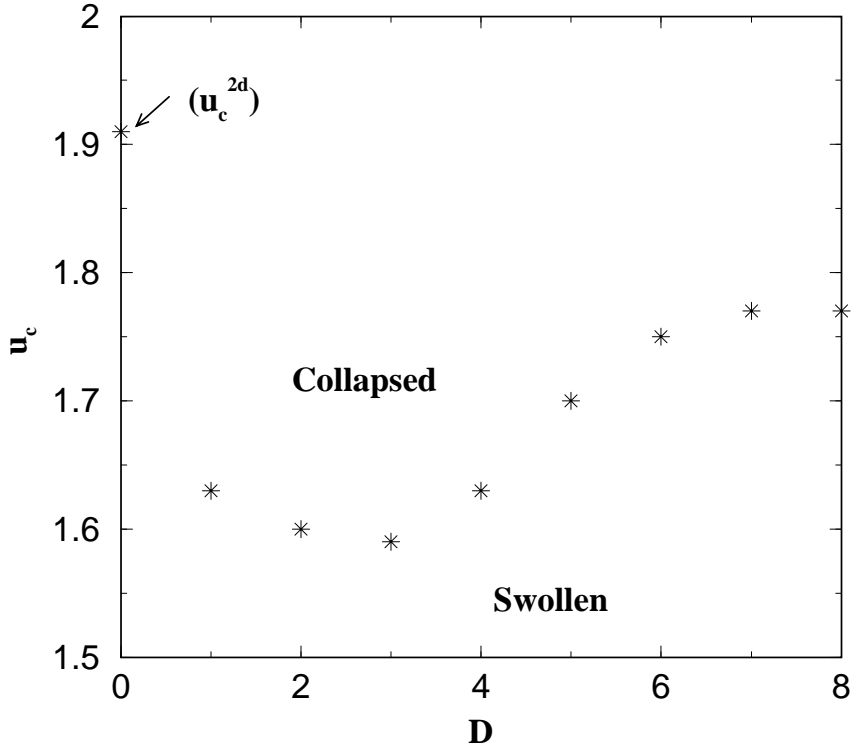}}}
\end{picture}
\caption[3]{The variation of $u$ with $D$ shows re-entrance
  behaviour in $3d$. The dip in $u$ indicates
  the rise in $\theta-$ temperature. It may be noted that
  this minimum value of $u_c$ shown in phase diagram is
  less than bulk ($u_c$=1.76 in $3d$) and surface ($u_c$=1.93
  in $2d$) value. }
\label{3}
\end{center}
\end{figure}
\begin{eqnarray}
G(D,u) = \lim_{N\rightarrow \infty} \frac{1}{N} \log Z_N(D,u)=\log[\mu(D,u)]. 
\label{@}\end{eqnarray}

To characterize conformational change in the system, we study specific
heat defined as $C_v=\partial^2 G/\partial \epsilon_p^2$ as a function of temperature. 
The peak in the $C_v-T$ diagram has
been identified as a sudden change in conformation.
Such changes in heat capacity in a narrow temperature range are due to
large fluctuations in energy. Studying the distribution of end to end distance in this
region we identify such transition as Coil -Globule transition. ofcourse, a sharp transition of a single chain occurs in the limit $N \to 
\infty$ only, and hence the accuracy of the extrapolation in Eq. 12 is
crucial. In the limit
$D \to \infty$, we find $u_c=1.93$ and $1.76$ for $2d$ and $3d$, respectively.
These values are in good agreement with earlier results$^{18-20}$. 
The variation of $u_c$, with $D$ has been shown in Fig. (6) which shows a 
re-entrance behaviour.
It can be seen from Fig. ($6$) that at a fixed   value of monomer-monomer attraction 
with a change of inter wall separation $D$, a polymer which was initially in 
the coil phase goes into the globule phase. With further increase 
of $D$ the polymer again attains the coil phase.
\section{Conclusions}
In this paper, we have studied the effect of confinement on the equilibrium 
thermodynamic properties of a linear polymer chain in various solvent
conditions.  The scaling relations found by us for $<R_{||}>$ and 
${<R_{\perp}}>$ are in agreement with earlier predictions$^2$. 
We show that due to the confinement, 
the end-to-end distance, $<R_E$> varies non-monotonically with inter- 
wall separation $D$. This is also reflected in variation of monomer density
with maxima at particular $ D^*$. In this region the number of non-bonded nearest
neighbours increases due to confinement and hence there is a rise in the transition 
temperature (i.e. decrease in $u_c$) as shown in Fig. (6).
\begin{figure}[htp!]
\begin{center}
\setlength{\unitlength}{1cm}
\begin{picture}(14,9)
%\put(-4,-.1){\scalebox{.9}[.9]{\includegraphics*{figure7.eps}}}
\put(-1.8,-.1){\scalebox{.9}[.9]{\includegraphics*{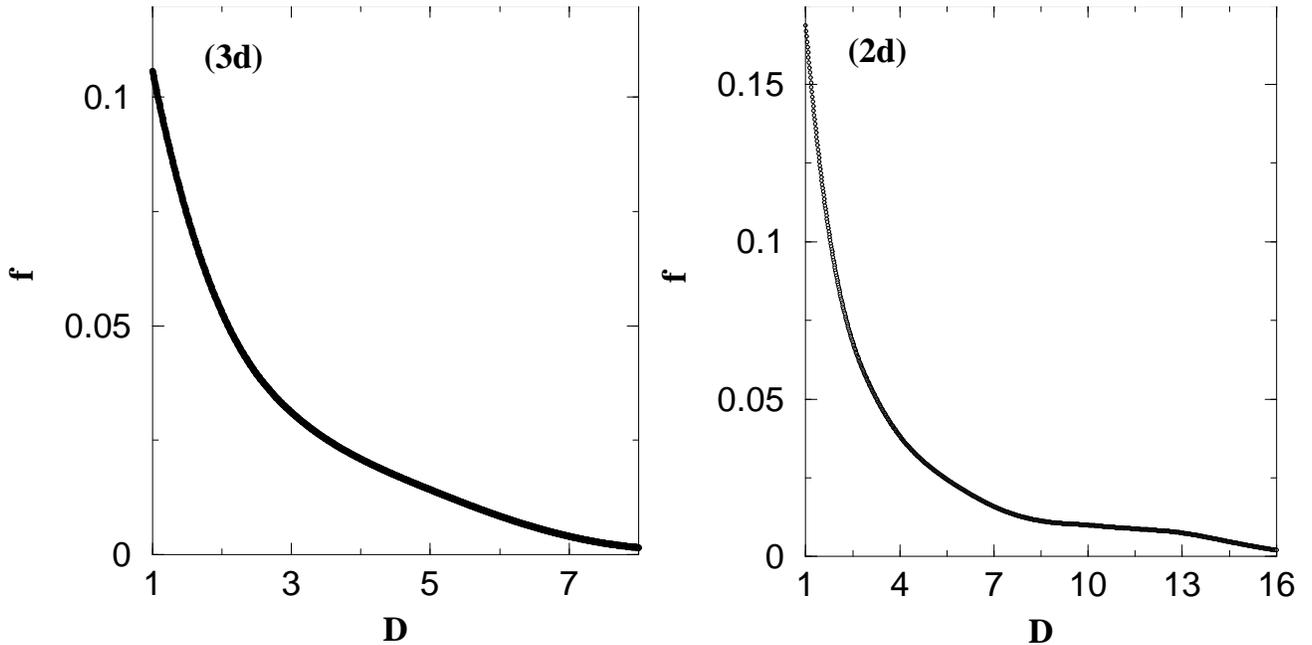}}}
\end{picture}
\caption[5]{ The variation of force exerted by confining
  walls on per monomer of the polymer chain with $D$. }
\label{5}
\end{center}
\end{figure}
The re-entrant behaviour can also be seen from Fig. (6). It has been observed that for a 
given value of monomer-monomer attraction a polymer which was initially 
in the coil phase goes into the globule phase due to the rise in nearest neighbour
pairs due to confinement at the particular $D^*$ . With further increase or decrease
 of the wall's 
separation $D$, the number of these pairs decreases and hence the polymer 
acquires the coil phase. 

We also studied this effect on a square lattice with finite $D$. We
do find a peak in the specific heat which signals that the system has
a tendency to be in the ordered state, but this tendency will only
turn into a phase transition when the slit width diverges
i.e. $D >  N^\nu$.

It is interesting to see the variation of force exerted by the confining wall
with $D$ in Fig. (7). It is qualitatively similar to the one obtained in a simulation$^{10}$,
in which the force has been applied by atomic force microscope tip on polymer chain
attached to the surface. In such a case it has been found that polymer
escapes from an approaching AFM tip (finite surface)by forming a stretched tether$^{10}$. 
Since in our work the surface has been considered as an infinite 
plane, "escape transition" is not possible. However, the work
with a finite surface of specific geometry is in progress.  

~\\Acknowledgment\\
%[-.9em]
We thank  Y. Singh for many fruitful discussions on the subject. 
One of us(SK) would like to acknowledge K. Binder
and W. Paul who suggested to analyze this problem through series analysis while he was visiting Condensed Matter Theory Group, Institute of Physics
, University of Mainz, Germany, supported by the DFG,SFB 625/A3. 
Financial support from CSIR(India) and, INSA (India) are gratefully acknowledged.\\
$^1$ P. Grassberger and H. P. Hsu, Eur. Phys. J. B {\bf 36} 209 (2003); 
P. Grassberger and H. P. Hsu, J. Chem. Phys., {\bf 120}, 2034 (2004).
\\
$^2$ A. Milchev and K. Binder, Eur. Phys. J. B {\bf 3}, 477 (1998); {\it ibid} {\bf 13} 607 
(2000); A. Milchev and K. Binder, J. Phys. II France  {\bf 6}, 21 (1996);
K. Binder, F. varnik, J. Baschnagel, P. Scheidler and W. Kob {\it Slow dynamics in 
Complex system: 3rd International Symposium}, edited by M. Tokuyama and 
I. Oppenheim,page 509 (2004).
\\
$^3$ M. Daoud, P.-G. de Gennes, J. Phys. {\bf 38}, 85 (1977); 
K. Kremer and K. Binder, J.Chem. Phys. {\bf 81} 638 (1984); I. Webman,
J. L. Lebowitz, M. H. Kalos,  J. Phys. II France  {\bf 41}, 579 (1980).
\\
$^4$ F. T. Wall, F Mandel and J. C. Chin, J. Chem. Phys. {\bf 65(6)}, 2231 (1976);
F. T. Wall and J. C. Chin, J. Chem. Phys. {\bf 66(7)}, 3066 (1977);
F. T. Wall, W. A. Seitz and J. C. Chin, J. Chem. Phys. {\bf 67(2)}, 434 (1977).
\\
$^5$ R. Barr, C. Brender and M. Lax, J. Chem. Phys. {\bf 72(4)}, 2702 (1980);
T. Ishinabe, J. Chem. Phys. {\bf 83(1)}, 423 (1985).
\\
$^6$ P. G. de Gennes, {\it Scaling concepts in polymer Physics}
(Carnell Univ. Press, Ithaca, 1979), des Cloiseaux J. and Jannink G. {\it Polymers in
Solution} (Oxford:Clarendon, 1990).
\\
$^7$ E. Eisenriegler , {\it Polymers Near Surfaces} (World Scientific, Singapore, 1993); 
G. J. Fleer, M. A. Cohen Stuart, J. M. H. M. Scheutjens, T. Cosgrove and B. Vincent, 
Polymers at Interfaces (Chapman and Hall, London, 1993).
\\
$^8$ T. M. Birshtein, E. B. Zhulina and A. M.
Skvortsov, Biopolymers {\bf 18}, 1171 (1979); G. J. Fleer and F. A. M. Leermakers, Curr. Opin.
Colloid Interface Sci. {\bf 2}, 308 (1997).
\\
$^9$ E. Eisenriegler, Phys. Rev. E {\bf 55}, 3116 (1997).
\\
$^{10}$ J. De Joannis, J. Jimenez, R. Rajagopalan and I. Bitsanis, Euro Phys. Lett. {\bf 51} 41 (2000).
\\
$^{11}$ R. Rajesh, D. Dhar, D. Giri, S. Kumar and Y. Singh, Phys. Rev. E
{\bf 65}, 056124 (2002); P. K. Mishra, D. Giri, S. Kumar and Y. Singh, Physica A {\bf 318}, 
171 (2003). 
\\
$^{12}$ S. Metzger, M. Muller, K. Binder and J. Baschnagel, J. Chem. Phys.
{\bf 118}, 8489 (2003).
\\
$^{13}$ Y. Singh, S. Kumar and D. Giri, J. Phys. A: Math.
Gen.{\bf 32}, L407 (1999); Y. Singh, D. Giri and S. Kumar, J. Phys. A: Math Gen.
{\bf 34}, L1 (2000); P. K. Mishra and Y. Singh, Phase Transitions:
{\bf 75 (4-5)}, 353 (2002).
\\
$^{14}$ A. J. Guttmann {\it Phase Transition and Critical Phenomena} edited by Domb C. and 
Lebowitz J. L. (Academic, New York, 1989) {\it Vol {\bf 13}}.
\\
$^{15}$ T. W. Burkhardt and I. Guim Phys. Rev. E {\bf 59} 5833 (1999).
\\
$^{16}$ G. A. Arteca, Macrmol. Theory Simul {\bf 8} 137 (1999); G. A. Arceta, Phys. rev. E {\bf 49} 2417 (1994) {\it ibid} {\bf 51} 2600 (1995).
\\
$^{17}$ J. H. van Vliet, M. C. Luyten and G. ten Brinke, Macromolecules {\bf 25} 3802 (1992).
\\
$^{18}$ P. Grassberger and R. Hegger Phys. Rev. E {\bf 51} 2674 (1995); P. Grassberger  and 
R. Hegger J. Physique I. France {\bf 5} 597 (1995).
\\
$^{19}$ R. Finsy, M. Janssens and A. Bellemanns J. Phys. A: Math. Gen. {\bf 8} L106 (1975).
\\
$^{20}$ D. P. Foster, E. Orlandini, M. C. Tesi J. Phys. A:Math. Gen. {\bf 25} L1211 (1992); 
D. P. Foster, J. M. Yeomans Physica A {\bf 177} 443 (1991).
\end{document}